\documentclass[12pt,hyperref]{article}  
\usepackage{graphicx}
\usepackage{fullpage}
\usepackage{epsfig}
\usepackage{color}
\usepackage{amsmath}
\usepackage{amssymb}

\begin{document}

\begin{center}
{\Large \bf Search for $Z_{c}^+(3900)$ in the $1^{+-}$ channel on the lattice }

\vspace{1cm}

Sasa Prelovsek$^{a,b}$ \footnote{e-mail: sasa.prelovsek@ijs.si} and Luka Leskovec$^{b}$

\vspace{0.3cm}

{\it a) Department of Physics, University of Ljubljana, 1000 Ljubljana, 
Slovenia} 

{\it b) Jozef Stefan Institute, Jamova 39, 1000 Ljubljana, Slovenia}

\end{center}

 \vspace{0.8cm}

\centerline{\large \bf Abstract}

\vspace{0.3cm}

Recently three experiments reported a discovery of manifestly exotic  $Z_c^+(3900)$ in the decay to $J/\psi\;\pi^+$, while $J$ and $P$ are experimentally unknown. We search for this state on the lattice by simulating the channel with $J^{PC}\!=\!1^{+-}$ and $I\!=\!1$, and we do not find a candidate for  $Z_c^+(3900)$. Instead, we only find discrete scattering states $D\bar D^*$ and $J/\psi\; \pi$, which inevitably have to be present in a dynamical QCD.  The possible reasons for not finding $Z_c^+$ may be that its quantum numbers are not $1^{+-}$ or that the employed interpolating fields are not diverse enough. Simulations with additional types of interpolators will be needed to reach a more definite conclusion.


\vspace{0.8cm}

\section{Introduction}
\label{sec:introduction}
Recently the BESIII collaboration \cite{Ablikim:2013mio} observed an interesting state $Z_{c}^+(3900)$ with mass $M_{Z_c}\!=\!3899.0\pm 3.6\pm 4.9~$MeV and $\Gamma_{Z_c}\!=\!46\pm10\pm20~$MeV, which decays to $J/\psi\;\pi^+$. This  suggests unconventional quark structure $\bar cc\bar du$. The discovery was confirmed by the Belle  \cite{Liu:2013dau} and CLEO-c  \cite{Xiao:2013iha} shortly after. While BESIII and Belle  reported only on the charged $Z_{c}^{\pm}(3900)$, the CLEO data also revealed a neutral $Z_{c}^{0}(3900)$ state. These three states are expected to form  an $I\!=\!1$ triplet and the neutral stat has $C=C_{J/\psi}C_{\pi^0}=-1$. The spin $J$ and parity $P$ have not been determined from the experiment yet. 

This is another in a series of unconventional states, which all happen to lie near a threshold. The state of interest lies close to the $D\bar D^*$ threshold and many theoretical studies suggest that the closeness of the threshold is related to its existence.  

Soon after discovery a number of theoretical works considered  this exotic state, employing effective field theories \cite{Wang:2013cya,Guo:2013sya,Li:2013xia}, 
QCD sum rules \cite{Chen:2010ze,Cui:2013yva, Zhang:2013aoa,Dias:2013xfa} and various different approaches  \cite{Voloshin:2013dpa,Braaten:2013boa,Wilbring:2013cha,
Chen:2013coa,Terasaki:2013lta, Liu:2013rxa,Wang:2013hga,Dong:2013iqa,Liu:2013vfa}.   

However, there was no first-principle lattice QCD simulation related to $Z_{c}^+(3900)$. The purpose of this paper is to make a first step along these lines. We perform a lattice QCD study using two flavors of dynamical quarks with $m_u=m_d$.   

The $J^P$ of $Z_{c}^+(3900)$ is experimentally not known and  most of phenomenological models listed above favor $J^P\!=\!1^+$, which corresponds to $J/\psi\; \pi$  or $D\bar D^*$ in $s$-wave. 
For this reason we consider the channel $J^{PC}\!=\!1^{+-}$ with $I\!=\!1$ in the present study, while other possible $J^P$ would have to be investigated in the future simulations. 

In a lattice QCD simulation, the states are identified from discrete energy-levels $E_n$ and in principle all 
 physical eigenstates with the  given quantum number appear. We employ  $J^{PC}\!=\!1^{+-}$,   $I\!=\!1$ and  total momentum zero. So the eigenstates are also the $s$-wave scattering states $D(\mathbf{p})\bar D^*(-\mathbf{p})$ and  $J/\psi(\mathbf{ p})\pi(-\mathbf{ p})$ with discrete momenta $\mathbf{p}$ due to periodic boundary conditions in space. If the two mesons do not interact then  $p\!=\!p^{n.i.}\!=\!\tfrac{2\pi}{L}|\mathbf{n}|$ and  the $M_1M_2$  scattering levels appear at   $E^{n.i.}=E_1(p^{n.i.})+E_2(p^{n.i.})$. In the presence of  interaction between $M_1$ and $M_2$, the scattering levels $E$ are shifted with respect to $E^{n.i.}$. The shift is negligible or small when the magnitude of the interaction between the two mesons is small.  Bound states and  resonances lead to levels in addition to the scattering levels. 

So  our major task is first to determine the discrete energy levels for $J^{PC}\!=\!1^{+-}$.  And then to find out whether  there are any extra energy levels in addition to the discrete scattering states $D\bar D^*$ and  $J/\psi\;\pi$. A signature for $Z_c^+(3900)$ would be an additional energy level near $E\simeq 3900~$MeV. 

 \section{Lattice setup}
 \label{sec:lat_set}

\begin{table}[t]
\begin{center}
\begin{tabular}{ccccc}
$N_L^3\times N_T$ & $a$[fm] & $L$[fm] & \#cfgs & $m_\pi~$[MeV]\\ 
\hline
$16^3\times32$ &  0.1239(13) & 1.98 & 280 &  266(4) \\
\end{tabular}
\caption{\label{tab:gauge_configs} Employed $N_f\!=\!2$ gauge configurations \cite{Lang:2011mn,Mohler:2012na}.}
\end{center}
\end{table}
 
This study is performed on one gauge ensemble kindly provided by the authors of \cite{Hasenfratz:2008ce,Hasenfratz:2008fg}. 
It is generated using a tree-level improved Wilson-Clover action with  dynamical $u/d$  quarks in the isospin limit $m_u=m_d$. The same action and the quark mass is used for valence $u/d$ quarks rendering $m_\pi=266(4)~$MeV.  Lattice parameters  are shown in Table \ref{tab:gauge_configs}, while further details are given in \cite{Lang:2011mn}. A rather small volume is favorable for the purpose at hand in order to avoid prohibitively dense $D\bar D^*$ scattering states, as explained in the next section. 
 The dynamical and valence quarks both obey periodic boundary conditions in space. We combine time periodic and anti-periodic valence perambulators into the 
so-called  $P+A$ perambulators in order to extend the effective time size of the lattice to $2N_T=64$ \cite{Lang:2011mn}.

To minimize the charm-quark discretization effects at a finite lattice spacing the Fermilab method \cite{ElKhadra:1996mp,Oktay:2008ex} is used for the charm quark. Its mass  is fixed by tuning the spin-averaged kinetic mass                        $\tfrac{1}{4}(M^{\eta_c}_2+3M^{J/\psi}_2)$ to its physical value \cite{Mohler:2012na}.  We used the same method on this ensemble and found good agreement with experiment for the conventional charmonium spectrum as well as masses and widths of charmed mesons in \cite{Mohler:2012na}, while a candidate for charmonium-like $X(3872)$ was found in \cite{Prelovsek:2013cra}. 

 According to the Fermilab method, the mass splittings between states involving a charm quark are expected to be close to the physical value even at finite lattice spacing. On the other hand, the rest masses are affected by large discretization effects. Consequently, we will quote energies $E-\tfrac{1}{4}(m_{\eta_c}+3m_{J/\psi})$ with respect to the spin-averaged mass  $\tfrac{1}{4}(m_{\eta_c}+3m_{J/\psi})$, where $am_{J/\psi}=1.54171(43)$ and $am_{\eta_c}=1.47392(31)$ on our lattice \cite{Mohler:2012na}.

\section{Energies of the non-interacting scattering states}

Since we will extract the energies of the interacting $D\bar D^*$ and $J/\psi\;\pi$  states, we first provide the energies of the single particles at  $p=0$ and $p=\tfrac{2\pi}{L}$, which are relevant in the energy region near $E\simeq 3900~$MeV. Those where extracted in \cite{Lang:2011mn,Mohler:2012na} and are collected in Table \ref{tab:single_particles}.

The non-interacting energies $E_{D}(p)\!+\!E_{D^*}(p)$  and $E_{J/\psi}(p)\!+\!E_{\pi}(p)$ of the scattering states are given by the dashed lines in Figs. \ref{fig:spectrum} and \ref{fig:spectrum_compare} for $p=0$ and $p=\tfrac{2\pi}{L}$. 
Note that  $D(0)\bar D^*(0)$ and $D(\tfrac{2\pi}{L})\bar D^*(-\tfrac{2\pi}{L})$ are separated by approximately $200~$MeV at our $L\simeq 2~$fm. In case of $L>2~$fm these scattering states would be separated by less than $200~$MeV and it would be more challenging to identify an additional state corresponding to $Z_c^+$. 

\begin{table}[t]
\begin{center}
\begin{tabular}{c|ccccc}
                        & $\pi$ & $J/\psi$ & $D$ & $D^*$  \\  
\hline
 $aE(p\!=\!0)$             & 0.1673(16) & 1.54171(43) & 0.9801(10) & 1.0629(13)       \\
$aE(p\!=\!\tfrac{2\pi}{L})$& 0.4374(64) & 1.5797(10)  & 1.0476(10) & 1.1225(14)        \\
\end{tabular}
\caption{\label{tab:single_particles} Energies $E(p)$ of single particles on our lattice for two relevant momenta $p$ \cite{Lang:2011mn,Mohler:2012na}.  }
\end{center}
\end{table}

\section{Interpolating fields}
\label{sec:interpolators}

We use six interpolators with $J^{PC}\!=\!1^{+-},\ I\!=\!1$ and total momentum zero. The interpolators below are written for  $Z_c^0$ with $I_3\!=\!0$ that has definite $C\!=\!-1$, but  the contractions and the resulting spectrum is exactly the same 
 for the charged $Z_c^{\pm}$ due to isospin symmetry $m_u\!=\!m_d$ in our simulation:  
\begin{align}
\label{interpolators}
O^{DD^*}_1&=[\bar c \gamma_5 u(p=0)~\bar u\gamma_i c(p=0) + \bar c \gamma_i u(p=0)~\bar u\gamma_5 c(p=0)] -\  \{u\to d\}\\
O^{DD^*}_2&=[\bar c \gamma_5 \gamma_t u(p=0)~\bar u\gamma_i \gamma_t c(p=0) + \bar c \gamma_i  \gamma_t u(p=0)~\bar u\gamma_5\gamma_t c(p=0)] -\  \{u\to d\}\nonumber\\
O^{DD^*}_3\!\!&=\!\!\!\!\sum_{p=\pm ~2\pi/L~e_{x,y,z}}[\bar c \gamma_5 u(p)~\bar u\gamma_i c(-p) + \bar c \gamma_i u(p)~\bar u\gamma_5 c(-p)] - \  \{u\to d\}\nonumber\\
O^{J/\psi\;\pi}_1&=\bar c \gamma_i c(p=0)~[\bar u\gamma_5 u(p=0)-\  \{u\to d\}]  \nonumber\\
O^{J/\psi\;\pi}_2&=\bar c \gamma_i \gamma_t c(p=0)~[\bar u\gamma_5 \gamma_t u(p=0)-\  \{u\to d\}]\nonumber\\
O^{J/\psi\;\pi}_3\!\!&=\!\!\!\! \sum_{p=\pm~ 2\pi/L~e_{x,y,z}}~\bar c \gamma_i c(p)~[\bar u\gamma_5 u(-p)-\  \{u\to d\}]~.\nonumber
\end{align}
All interpolators are of meson-meson type, where 
momenta is projected separately for each meson current: $\bar q_1\Gamma q_2(p)\equiv  \sum_{x}e^{ipx}\sum_{c=1,2,3}~q_1^c(x,t)\Gamma q_2^c(x,t)$.
They transform according to the irreducible representation $T_1^{+-}$ of the discrete lattice symmetry group $O_h$, which contains $J^{PC}\!=\!1^{+-}$; it  contains in principle also $J^{PC}\geq 3^{+-}$ states, but those are above the region of interest. 
 All quark fields  are smeared  according to the Laplacian Heaviside smearing $q\equiv \sum_{k=1}^{N_v}v^{(k)}v^{(k)\dagger}q_{point}$ \cite{Peardon:2009gh,Mohler:2012na} with $N_v\!=\!96$ for $O^{DD^*,J/\psi\pi}_1$ and $N_v\!=\!64$ for the remaining four. 

Although the interpolators (\ref{interpolators}) have a form, which gives significant coupling to the $D\bar D^*$ and $J/\psi\;\pi$ eigenstates, we emphasize that they in principle couple to all physical eigenstates with a given quantum number.  We would expect that six interpolators (\ref{interpolators})  are sufficiently many and also sufficiently linearly independent that some of them would couple also to $Z_c$ if it has $J^{PC}\!=\!1^{+-}$.  

Note that all interpolators in (\ref{interpolators}) are products of two color-singlet currents. Interpolators with diquark anti-diquark color structure $[qq]^{\bar 3_c}~[\bar q\bar q]^{3_c}$ are not implemented explicitly. However  this structure is formally  present in a linear combination after interpolators  (\ref{interpolators}) are written as a sum according to the  Fierz transformations.

\section{Calculation of the energy levels and overlaps}

The energies $E_n$ of the eigenstates $|n\rangle$ are extracted from the time-dependence of the $6\times 6$ correlation matrix 
\begin{equation}
\label{c}
C_{ij}(t)= \langle 0| O_i (t+t_{src})~ O^\dagger_j(t_{src})|0\rangle=\sum_{n}Z_i^nZ_j^{n*}~e^{-E_n t}\qquad Z_i^n\equiv \langle O_i|n\rangle
\end{equation}
for every $t$ and every second $t_{src}$.

\begin{figure*}[t]
\begin{center}
\includegraphics*[width=0.44\textwidth,clip]{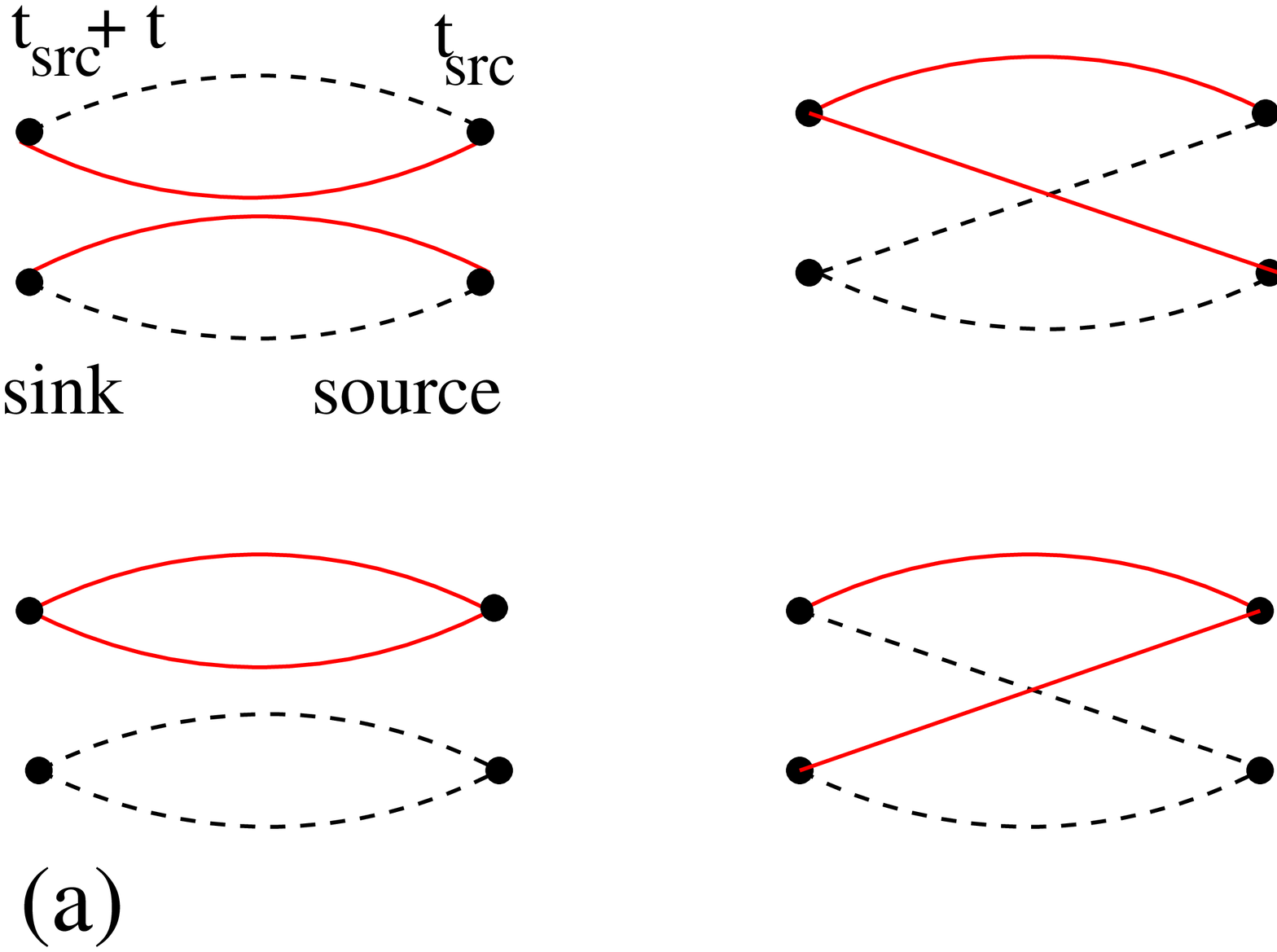}$\qquad$$\qquad$
\includegraphics*[width=0.44\textwidth,clip]{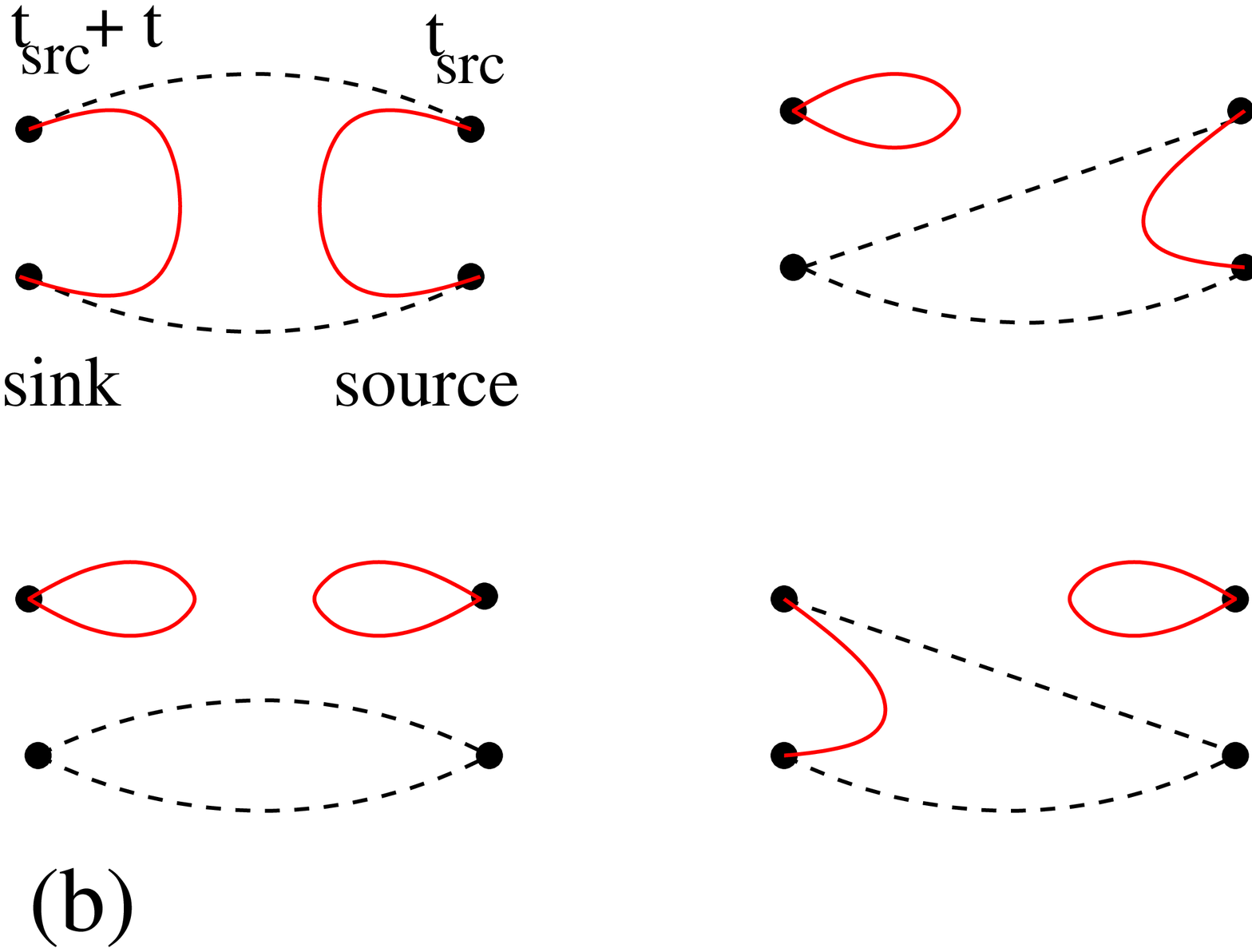} 
\end{center}
\caption{\label{fig:contractions}Wick contractions that enter correlation matrix (\ref{c}) for interpolators (\ref{interpolators}) with $I\!=\!1$: the solid red lines represent $c$ quark, while the dashed black lines represent $u$ or $d$ quark.  All contractions are calculated, but we present results based only on the contractions in Fig. (a), where the charm quark propagates from the source to sink.  }
\end{figure*}

We calculate all the  Wick contractions in Figs. \ref{fig:contractions}a and \ref{fig:contractions}b entering  the $6\times 6$  correlation matrix $C_{ij}(t)$. They are evaluated    using the distillation method \cite{Peardon:2009gh}. The contractions in Fig. \ref{fig:contractions}b involve charm annihilation  and the correlation functions are dominated by the propagation of the light quarks. Their effect on  charmonium states  is  suppressed due to the Okubo-Zweig-Iizuka rule, it was explicitly verified to be very small for the conventional charmonium \cite{Levkova:2010ft} and we postpone the study of their effects to a future publication. Note that almost all previous simulations of charmonium or charmonium-like states omitted charm-annihilation diagrams. 
 In the present paper we present results, where $C_{ij}(t)$ contains contractions in Figs. \ref{fig:contractions}a, where the charm quarks propagate from source to sink.

\begin{figure*}[t]
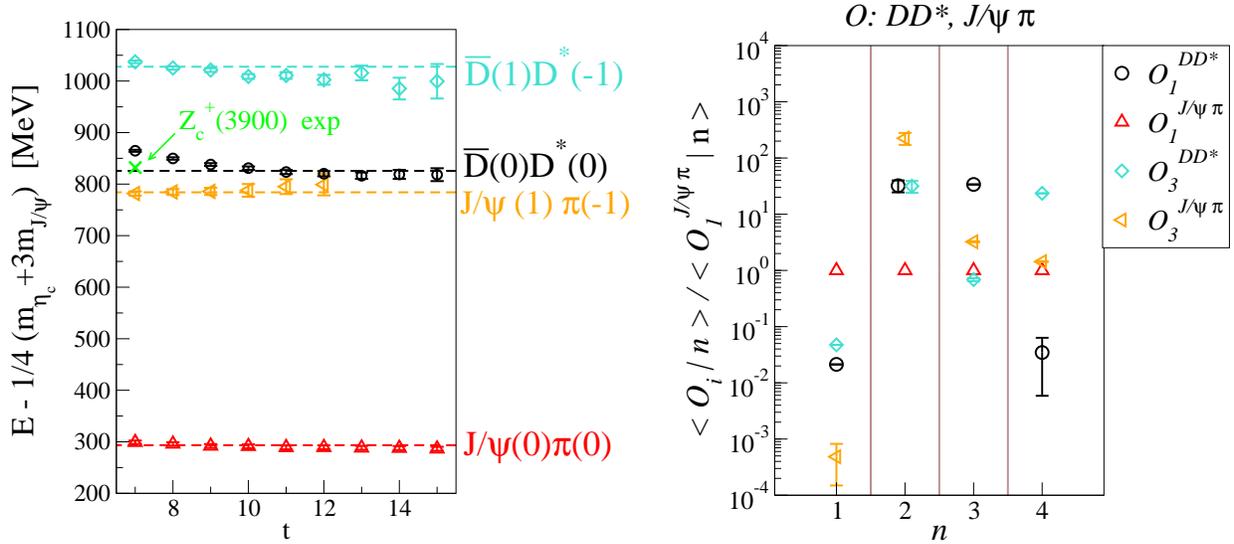

\begin{center}
\includegraphics*[width=0.5\textwidth,clip]{delMeV_T1pm.eps}$\quad$
 \includegraphics*[width=0.45\textwidth,clip]{rat0_T1pm.eps} 
\end{center}
\caption{\label{fig:spectrum}Left: The energy spectrum in the $J^{PC}\!=\!1^{+-}$ channel with $I\!=\!1$.  Effective energies $E_{n}^{eff}(t)-\tfrac{1}{4}(m_{\eta_c}+3m_{J/\psi})$ correspond to the spectrum $E_{n}-\tfrac{1}{4}(m_{\eta_c}+3m_{J/\psi})$ in the plateau region. The dashed lines correspond to the energies of the non-interacting scattering states indicated on the right, which are also extracted from our lattice. Right: Overlaps $\langle O_i|n\rangle$ of eigenstates $n=1,..,4$ from the left plot to our interpolators $O_i$ (\ref{interpolators}). Overlaps are presented in a form of a ratio with respect to  $\langle O_1^{J/\psi\; \pi}|n\rangle$. }
\end{figure*}

  To recover as many energy levels as possible, currently the best method is the variational method
 \cite{Luscher:1990ck,Michael:1985ne,Blossier:2009kd} also known as the generalized eigenvalue problem
\begin{align}
C(t)\vec{v}_{n}(t)=\lambda_{n}(t)C(t_{0})\vec{v}_{n}(t)~.
\end{align}
Our results are consistent for $2\leq t_0\leq 4$ and we present them for $t_0\!=\!2$. The time dependence of the  eigenvalues $\lambda_{n}(t) \rightarrow e^{-E_{n}(t-t_{0})}$ gives the energies $E_n$ of the eigenstates. These will be presented in Figs. \ref{fig:spectrum} and \ref{fig:spectrum_compare} by means of the  effective energies $E_n^{eff}(t)$, which equal $E_n$ in the plateau region
\begin{equation}
E_n^{eff}(t)\equiv \log \frac{\lambda_n(t)}{\lambda_n(t+1)}\to E_n~. 
\end{equation}

The composition of the resulting eigenstates $|n\rangle$ will be illustrated in terms of their overlaps $\langle O_i|n\rangle$ with the employed interpolators. For this purpose we evaluate the ratios of overlaps for the state $n$ to two different interpolators \cite{Blossier:2009kd}
\begin{equation}
 \frac{\langle {\cal O}_i|n\rangle}{\langle {\cal O}_j|n\rangle}=\frac{\sum_k C_{ik}(t)u_k^n(t)}{\sum_{k'} C_{jk'}(t)u_{k'}^n(t)}~.
 \end{equation} 
These are indeed almost independent of $t$ and they are evaluated at $t=8$ in  Fig. \ref{fig:spectrum}.  

\section{Results for $J^{PC}=1^{+-}$ and $I=1$}
\label{sec:results}

The main result of our study are the discrete energy levels $E_n$ plotted in Fig. \ref{fig:spectrum}. Instead of $E_n$, we plot $E_n^{eff}(t)-\tfrac{1}{4}(m_{\eta_c}+3m_{J/\psi})$, where the dominant discretization errors related to the charm quark cancel. The time-dependence shows rather good plateaus, therefore 
$E_n^{eff}(t)-\tfrac{1}{4}(m_{\eta_c}+3m_{J/\psi})$ represents $E_n-\tfrac{1}{4}(m_{\eta_c}+3m_{J/\psi})$ in the plateau region. The dashed lines represent the energies of the non-interacting $D\bar D^*$ and $J/\psi \pi$ scattering states. 

In the energy region of interest, we observe four energy levels and  all of them   almost exactly coincide with the energies of the   non-interacting scattering states $J/\psi(0)\pi(0)$, $J/\psi(\tfrac{2\pi}{L})\pi(-\tfrac{2\pi}{L})$, $D(0)\bar D^*(0)$ and  $D(\tfrac{2\pi}{L})\bar D^*(-\tfrac{2\pi}{L})$. We conclude that the observed levels correspond to these scattering states and that the magnitude of the interaction between $J/\psi$ and $\pi$ or between $D$ and $\bar D^*$ is small.   

The main conclusion based on  Figure \ref{fig:spectrum} is that we do not find any additional energy level which could be related to  $Z_c^+$. Experimental $Z_c^+(3900)$ is found at the energy  represented by the green cross, which is close to the $DD^*$ threshold. Although we find a nearby energy level at $E_3-\tfrac{1}{4}(m_{\eta_c}+3m_{J/\psi})\simeq 820~$MeV, represented by the black circles, we believe it is related to the $D(0)\bar D^*(0)$ scattering state, which inevitably has to appear as an energy level in dynamical QCD.  

The right pane in the Figure \ref{fig:spectrum} indicates that the levels $n=1,2,3,4$ have largest overlaps with $O_1^{J/\psi\;\pi}$, $O_3^{J/\psi\;\pi}$, $O_1^{DD^*}$ and $O_3^{DD^*}$, respectively. This is another signature that they correspond to the almost non-interacting scattering states. 

We conclude that we do not find a candidate for $Z_c^+(3900)$ in the channel with $J^{PC}\!=\!1^{+-}$ and $I\!=\!1$ in our lattice simulation with degenerate dynamical $u/d$ quarks and $m_\pi\!\simeq\! 266(4)~$MeV. 

\section{Discussion and  outlook}

In the following, several possibilities are listed which might be responsible that we do not find a candidate for $Z_c^+(3900)$ in the channel with $J^{PC}\!=\!1^{+-}$ and $I\!=\!1$:
\begin{itemize}
\item The $J^P$ of the discovered $Z_c^+(3900)$ is experimentally not known and it may be that $J^P\not = 1^+$. A future  simulation for other $J^P$ is required  to investigate this.  
\item The experimentally  observed enhancement in the $J/\psi\; \pi$ invariant mass near $\sqrt{s}\simeq 3900~$MeV might be due to some non-conventional $D\bar D^*$ threshold effect rather than the genuine resonance. Since the existence of $Z_c^+(3900)$ was confirmed by three experiments, we consider that as a less likely possibility. 
\item In  case that $Z_c^+(3900)$ with $J^{PC}\!=\!1^{+-}$ and $I\!=\!1$ exists in nature, it is still possible that the employed interpolators (\ref{interpolators}) are not diverse enough and consequently might  not lead to an additional level related to it. As discussed in Section \ref{sec:interpolators}, the interpolators (\ref{interpolators}) are expected to have particularly good coupling to the scattering states. But these interpolating fields do not correspond to the true eigenstates, so they are expected to couple in principle to all physical eigenstates (including possible $Z_c^+$) with a given quantum number. In practice, they might  not be diverse enough to render exotic $Z_c^+$ in addition to all the nearby scattering states.

We note that in analogous simulation of the $J^{PC}\!=\!1^{++}$ channel with $I=0$, the $J/\psi\;\omega$ and $D\bar D^*$ interpolators alone lead to all the scattering states and the $\chi_{c0}(1P)$ state, but did not render $X(3872)$ near $D\bar D^*$ threshold (see Fig. 1d in \cite{Prelovsek:2013cra}). Only when meson-meson interpolators were combined with the  $\bar cc$ interpolators,   an energy level related to $X(3872)$ appeared. The $\bar cc$ interpolators can obviously be of no help for the manifestly exotic channel considered in this paper. 

To make a more definite conclusion, we suggest to perform  a simulation including meson-meson interpolators (\ref{interpolators}) as well as some different type of interpolators that are not products of two color-singlet currents. Valuable color structures would be diquark anti-diquark interpolators such as $[qq]^{\bar{3}_c} [\bar{q}\bar{q}]^{3_c}$ and $[qq]^{6_c} [\bar{q}\bar{q}]^{\bar{6}_c}$ or the color octet combination $[q\bar{q}]^{8_c}[q\bar{q}]^{8_c}$. These also allow for several combinations in the Dirac space, which lead to the desired $J^{PC}$. The energy levels would need to be calculated using the combined interpolator basis. A signature in favor of $Z_{c}(3900)$ would be an energy level with $E \approx 3900$ MeV in addition to the $D \bar{D}^*$ and $J/\psi \pi$ discrete scattering levels.

\item The non-observation of $Z_c^+$ could also be  related to unphysically high $m_\pi\!\simeq\! 266~$MeV or the  exact isospin $m_u=m_d$ in our simulation, but we consider these as unlikely possibilities.   
\end{itemize}

\begin{figure*}[htb]
\begin{center}
\includegraphics*[width=0.85\textwidth,clip]{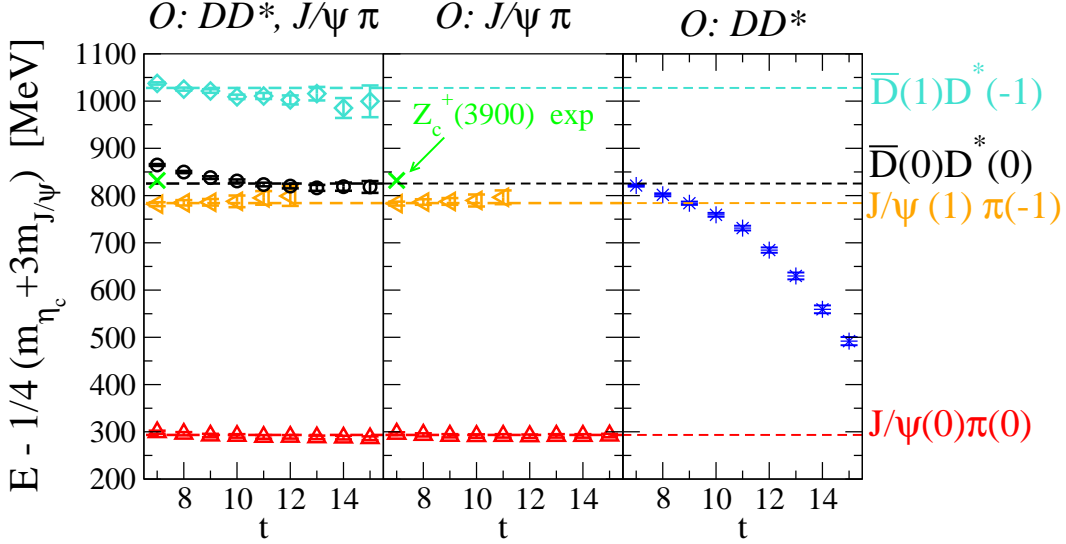} 
\end{center}
\caption{\label{fig:spectrum_compare}  The energy spectrum in the $J^{PC}\!=\!1^{+-}$ channel with $I\!=\!1$ for three choices of correlation matrices. The left plot shows the result including all interpolators in (\ref{interpolators}). The middle plot results from using just $3\times 3$ sub-matrix with $O_{1,2,3}^{J/\psi\; \pi}$ interpolators, while the right plot shows the ground state based on  $3\times 3$ sub-matrix with $O_{1,2,3}^{D\bar D^*}$ interpolators. }
\end{figure*}

Our result is based on a single ensemble, but we believe that the continuum limit $a\to 0$ does  not modify our conclusion regarding the non-observation of the level related to $Z_c^+$. Infinite volume limit is more tricky in this respect as there would be a continuum of scattering states. We expect that a rather small volume $L\simeq 2~$fm  still allows for establishing an existence of an addition level while this strategy becomes more challenging for significantly larger volumes.  

In order to provide some guidance for the future simulations aiming at this interesting state, we present energy levels obtained using only $O_{1,2,3}^{J/\psi\; \pi}$ or only $O_{1,2,3}^{DD^*}$  in Fig. \ref{fig:spectrum_compare}. This Figure indicates that one type alone can not lead to the final conclusion. The $O^{J/\psi\; \pi}$ interpolators alone give $J/\psi\; \pi$ scattering states, but can not conclude on the existence of $Z_c^+$ which is located near the $DD^*$ threshold. Also $O^{D\bar D^*}$ alone do not help as they couple to the $J/\psi(0)\pi(0)$ ground state, which is manifested by the falling effective energy in  Fig. \ref{fig:spectrum_compare}.  Therefore future simulations need to consider $J/\psi\;\pi$ as well as $D\bar D^*$ interpolating fields, preferably combined with yet another type.

\section{Conclusions}

We do not find a candidate for $Z_c^+(3900)$ in the channel with $J^{PC}\!=\!1^{+-}$ and $I\!=\!1$ in our lattice simulation with degenerate dynamical $u/d$ quarks and $m_\pi\!\simeq\! 266(4)~$MeV. This conclusion is based on the simulation of this channel using $J/\psi\;\pi$ and $D\bar D^*$  interpolating fields, where only discrete $J/\psi\;\pi$ and $D\bar D^*$ scattering states are found but no additional candidate for $Z_c^+(3900)$. 
To make a final conclusion regarding the existence of $Z_c^+(3900)$ in the $J^{PC}\!=\!1^{+-}$ channel, we propose a  future simulation including $J/\psi\;\pi$ and $D\bar D^*$  interpolators as well interpolators of yet another type.

\vspace{1cm}

{\bf Acknowledgments}

We thank Daniel Mohler for pointing out the experimental discovery of $Z_c^+(3900)$.   
We  acknowledge Anna Hasenfratz for providing the gauge configurations, as well as  Daniel Mohler and C. B. Lang for providing the perambulators. This work is supported by the Slovenian Research Agency ARRS and is a preparatory work for the ARRS project  N1-0020 and FWF project I1313-N27. 

\clearpage

\bibliographystyle{h-physrev4}
\bibliography{refs_Zc3900}

\end{document}